\documentstyle[aps, prbbib, tighten, multicol]{revtex}

\begin{document}
\draft

\title{Electronic transport through a local state \\linearly coupled to phonon modes}
\author{M. A. Gata\cite{MAG}}

\address{Department of Physics,\\The University of Texas at Austin,\\ Austin, Texas 78712\\and\\
DCTD, University of the Azores,\\ 9500 Ponta Delgada\\ Azores, Portugal}

\maketitle

\begin{abstract} 
A resolvent formalism is applied to the problem of inelastic scattering of an electron linearly coupled to a set of phonon
modes. It is shown how the many phonon mode coupling and excitation can be reduced to a single phonon mode description,
consistent with other approaches to the phenomenon of inelastic tunneling in heterostructures. A unifying connection is made
with the phenomena of resonant electron scattering through molecules in the gas phase and of tunneling of electrons
through adsorbates in scanning tunneling microscopy.
\end{abstract}
%
%
%
%%%%%%%%%%%%%%%%%%%%%%%%%%%%%%%%%%%%%%%%%        I. Introduction           %%%%%%%%%%%%%%%%%%%%%%%%%%%%%%%%%%%%%%%
%
\section{Introduction}

Recently, we addressed the problem of the correspondence between two different descriptions of the phenomenon of inelastic
electronic transitions between two distinct set of continuum electron states, through an intermediate resonant state coupled to
an oscillator \cite{G+A99}. Even though the problem of interest there was the influence of vibration damping on the
spectroscopy of adsorbates with the scanning tunneling microscope, it was shown that the formalism adopted, taken over from a
description of resonant tunneling in semiconductor structures \cite{Wingreen} with the necessary modifications, was in fact
reducible to a simpler resolvent method, in the no-damping limit, but with the restriction of just one vibration coordinate.
Our intent at that point was to make a connection with a previous paper \cite{G+A93} on STM, but an obvious possibility then
 was the eventual extension of that correspondence between the two approaches,
from just one vibration coordinate to a Brioullin zone of modes. The present paper addresses this issue, showing that indeed a
simple resolvent model, without appealing to many-body field theory, seems to be able to describe inelastic tunneling in small
semiconductor structures, at least with the restrictions of a $0$ K temperature and an Einstein band of optical phonons. Even
so, we believe it is of some interest to present this analysis, as it may provide a simpler means of assessing the essential
physics of the situation and also as a starting point for further useful extensions.

On the other hand, even though this model can be applied to many real distinct situations in physics and in chemistry, as
pointed out by Gadzuk\cite{Gadzuk91} when drawing similar comparisons, we will be referring specifically to inelastic resonant
electronic transport in semiconductors (for a recent exposition based on nonequilibrium Green functions, see Ref.\
\onlinecite{Haug+Jauho}), so that our results can be compared directly with the more general, and more powerful, many-body
approach used in Ref.\ \onlinecite{Wingreen}, which employs a two-particle many-body Green's function and which was
subsequently simplified to a singe-particle many-body Green's function treatment \cite{Haug+Jauho}. Other methods
have been published, such as those of Ref.\ \onlinecite{G+S}, for example. But with the present, simpler, method even the
inclusion of more than one intermediate, resonant state should not introduce major difficulties. The essentials of this method
were used many years ago by Domcke and Cederbaum \cite{D+C}, to describe inelastic electron scattering by molecules in the gas
phase.

%
%
%%%%%%%%%%%%%%%%%%%%%%%%%%%%%%        II. Resonance Green's Phonon Operator        %%%%%%%%%%%%%%%%%%%%%%%%%%%%%%%%%%%%
%
\section{Resonance Green's Phonon Operator}

As usual, the total spinless Hamiltonian is a sum of three parts, the electron Hamiltonian, the phonon Hamiltonian and the
electron-phonon interaction, $H=H_{el}+H_{ph}+H_{int}$:
\begin{eqnarray}
H_{el}  =& &
\epsilon^{0}_{a} c^{\dagger}_a c_a +\sum_{l}\epsilon_l c^{\dagger}_l c_l +\sum_{k}\epsilon_k c^{\dagger}_kc_k \nonumber \\
&+& \sum_{l}(V_{al} c^{\dagger}_a c_l+V_{la}c^{\dagger}_l c_a)\nonumber\\
&+& \sum_{k}(V_{ak} c^{\dagger}_a c_k+V_{ka}c^{\dagger}_k c_a),
\end{eqnarray}
\begin{equation}
H_{ph}=\sum_{q}\hbar \omega_{q} (b_{q}^{\dagger}b_{q}+1/2), 
\end{equation}
and
\begin{equation}
H_{int} =c^{\dagger}_a c_a \sum_{q} M_{q}(b_{q}^{\dagger} + b_{-q}).
\end{equation}
In $H_{el}$ the first term describes the localized resonant electron state, the second and third terms describe the electronic
states in the left and right (isolated) leads, and the last two terms represent the hybridization between the localized
electronic resonance and the two leads. $H_{ph}$ describes the harmonic phonon states and $H_{int}$ is the
linear electron-phonon interaction, whose strength is given by the coupling constant $M_{q}$, acting only when an electron
occupies the intermediate, resonant state.

We proceed to compute the resonance Green's function from the corresponding resolvent operator $G=(\epsilon - H)^{-1}$, using
the equations-of-motion method, as in the original Anderson description of localized magnetic states associated with impurities
in metals \cite{Anderson61,Jones+March2}. In doing this, we assume the validity of the Born-Oppenheimer approximation and
include the interaction Hamiltonian in the resonant electron state energy, writing it as a phonon operator
\begin{equation}
\epsilon_{a} = \epsilon_{a}(b^{(\dagger)}_{q}) =\epsilon^{0}_{a} + \sum_{q} M_{q}(b_{q}^{\dagger} + b_{-q}),
\end{equation}
thereby rendering the electron Hamiltonian $H_{el}$ adiabatically dependent on the state of the lattice oscillators.

From the equations of motion for the operators
\begin{equation}
i\hbar\frac{\partial c_{\alpha}}{\partial t} = [c_{\alpha}, H]
\end{equation}
we obtain, taking into account the commutation of electron and phonon operators under the adiabatic hypothesis, the set of
equations
\begin{eqnarray}
i\hbar\frac{\partial c_{a}}{\partial t} =& &\epsilon_{a} c_{a}(t) + \sum_{k} V_{ak} c_{k}(t) + \sum_{l} V_{al}
c_{l}(t)\nonumber\\ i\hbar\frac{\partial c_{k}}{\partial t} =& &\epsilon_{k} c_{k}(t) + V_{ka} c_{a} (t)\nonumber\\
i\hbar\frac{\partial c_{l}}{\partial t} =& &\epsilon_{l} c_{l}(t) + V_{la} c_{a} (t).
\end{eqnarray}
Fourier transforming to the energy domain, we get the corresponding algebraic system
\begin{eqnarray}
\epsilon c_{a}(\epsilon) =& & \epsilon_{a} c_{a}(\epsilon) + \sum_{k} V_{ak} c_{k}(\epsilon) + \sum_{l} V_{al} c_{l}(\epsilon)
\nonumber\\
\epsilon c_{k}(\epsilon) =& & \epsilon_{k} c_{k}(\epsilon) + V_{ka} c_{a}(\epsilon) , \forall k,\nonumber\\
\epsilon c_{l}(\epsilon) =& & \epsilon_{l} c_{l}(\epsilon) + V_{la} c_{a}(\epsilon) , \forall l, 
\end{eqnarray}
which can be rewritten in matrix form
\begin{eqnarray}
\left[ \epsilon {\bf 1} - \left( \begin{array}{c}\epsilon_{a} \\ {\bf V_{ka}} \\ {\bf V_{la}}\end{array}
                                 \begin{array}{c} {\bf V_{ak}^{T}} \\ {\bf\epsilon_{k}} \\ {\bf 0}\end{array}
                                 \begin{array}{c} {\bf V_{al}^{T}} \\ {\bf 0} \\ {\bf\epsilon_{l}}\end{array}\right)\right]
\otimes
                                 \left(\begin{array}{c} c_{a} \\ {\bf c_{k}} \\ {\bf c_{l}} \end{array} \right) = 
                                 \left(\begin{array}{c}  0   \\ {\bf 0} \\ {\bf 0} \end{array} \right),
\end{eqnarray}
where ${\bf V_{ka}}$ is a column matrix, ${\bf V_{ak}^{T}}$ the row matrix transpose of  ${\bf V_{ak}}$,
${\bf\epsilon_{k}}$ and ${\bf\epsilon_{l}}$ are diagonal matrices with elements $\epsilon_{k_1}, \epsilon_{k_2}, \cdots$ and
$\epsilon_{l_1}, \epsilon_{l_2}, \cdots$, respectively, ${\bf 1}$ is the identity matrix of the appropriate dimension and ${\bf
0}$ are null matrices of the appropriate dimension also.

This matrix equality can be put in the form
\begin{equation}
\left(\epsilon {\bf 1} - {\bf F_{el}}\right)\otimes{\bf c} = {\bf 0},
\end{equation}
writing the previous matrices more compactly, in an obvious notation. Introducing now another diagonal phonon operator matrix
\begin{equation}
{\bf F_{ph}}=\left[\sum\hbar\omega_{q}(b^{\dagger}_{q}b_{q} + 1/2)\right]{\bf 1},
\end{equation}
we define an electron-phonon resolvent operator 
\begin{eqnarray}
{\bf G^{total}}(\zeta) & = & \left[(\zeta + i\eta){\bf 1}-({\bf F_{el}}+{\bf F_{ph}})\right]^{-1}\nonumber\\
                       & = & \left[(\zeta + i\eta){\bf 1}-{\bf F^{total}}\right]^{-1},
\end{eqnarray}
$\zeta$ being the total energy parameter and $\eta$ a positive infinitesimal. Then $\left[(\zeta + i\eta){\bf 1}-
{\bf F^{total}}\right]\otimes{\bf G^{total}}(\zeta) ={\bf1}$ which, expanded, means
\begin{eqnarray}
\left(\begin{array}{c}(\zeta+i\eta) - \epsilon_{a} - H_{ph}{\bf 1} \\ {\bf -V_{ka}} \\ {\bf -V_{la}}\end{array}
      \begin{array}{c} {\bf -V_{ak}^{T}} \\ (\zeta+i\eta){\bf 1}-{\bf\epsilon_{k}}- H_{ph}{\bf 1} \\ {\bf 0}\end{array}
      \begin{array}{c} {\bf -V_{al}^{T}} \\ {\bf 0} \\ (\zeta+i\eta){\bf 1}-{\bf\epsilon_{l}}- H_{ph}{\bf1}\end{array}\right)
\nonumber\\
\otimes
       \left(\begin{array}{c}    G_{aa}    \\ {\bf G_{ka}} \\ {\bf G_{la}} \end{array}
             \begin{array}{c} {\bf G_{ak}} \\ {\bf G_{kk}} \\ {\bf G_{lk}} \end{array}
             \begin{array}{c} {\bf G_{al}} \\ {\bf G_{kl}} \\ {\bf G_{ll}} \end{array} \right)
= 
       \left(\begin{array}{c}    1      \\ {\bf 0}      \\ {\bf 0}      \end{array}
             \begin{array}{c} {\bf 0}   \\ {\bf 1}      \\ {\bf 0}      \end{array}
             \begin{array}{c} {\bf 0}   \\ {\bf 0}      \\ {\bf 1}      \end{array}\right).
\end{eqnarray}
Introducing a common index $m$ denoting both $k$ and $l$, we have the system of equations
\begin{eqnarray}
[(\zeta+i\eta) - \epsilon_{a} - H_{ph}] G_{aa} - \sum_{m=k,l}V_{am}G_{ma} = 1\nonumber\\
-V_{ma} G_{aa} + [(\zeta+i\eta) - \epsilon_{m} - H_{ph}] G_{ma} = 0  .
\end{eqnarray}
In the second equalitiy we solve for $G_{ma}$ (where $m=k,l$) and substitute it in the first, obtaining the Green's phonon
operator for the localized electronic resonance
\begin{eqnarray}
G_{aa}(\zeta)=& &\{[(\zeta+i\eta) - \epsilon_{a} - H_{ph}]\nonumber\\
&-&\sum_{m=k,l}\left|V_{am}\right|^{2}[(\zeta+i\eta)-\epsilon_{m}-H_{ph}]^{-1}\}^{-1}
\end{eqnarray}
as a function of the total energy parameter, $\zeta$. At zero K temperature, the initial energy in the transition process, equal
to the total energy, will comprise the incoming electron energy, $\epsilon_{k_i}$ plus the ground state lattice energy,
$1/2 \sum_{q} \hbar\omega_{q}$. With this last quantity fixed, we use as the energy parameter the incoming electron energy and
write
\begin{eqnarray}
G_{aa}(\epsilon_{k_i})=\{[(\epsilon_{k_i}+1/2\sum_{q} \hbar\omega_{q} + i\eta) - \epsilon^{0}_{a} -
\sum_{q}M_{q}(b^{\dagger}_{q}+b_{-q}) - \sum_{q}\hbar\omega_{q} (b^{\dagger}_{q}b_{q}+1/2)]\nonumber\\
- \sum_{m}\left|V_{am}\right|^{2}[\epsilon_{k_i}+1/2\sum_{q}\hbar\omega_{q}+i\eta-\epsilon_{m}
-\sum_{q}\hbar\omega_{q}(b^{\dagger}_{q}b_{q}+1/2)]^{-1}\}.
\end{eqnarray}
However, since in general $(x+i\eta)^{-1}=P(x^{-1})-i\pi\delta(x)$, $P$ designating the principal part, we get
\begin{eqnarray}
[\epsilon_{k_i}+i\eta-\epsilon_{m}-\sum_{q}\hbar\omega_{q}b^{\dagger}_{q}b_{q}]^{-1}\nonumber\\
= P[(\epsilon_{k_i}-\epsilon_{m}-\sum_{q}\hbar\omega_{q}b^{\dagger}_{q}b_{q})^{-1}]\nonumber\\
-i\pi\delta(\epsilon_{k_i}-\epsilon_{m}-\sum_{q}\hbar\omega_{q}b^{\dagger}_{q}b_{q})
\end{eqnarray}
and defining an electronic shift $\Delta$ and width $\Gamma$ phonon operators by
\begin{eqnarray}
\Delta (\epsilon_{k_i}) = P \sum_{m=k,l}\left|V_{am}\right|^{2}
                           (\epsilon_{k_i}-\epsilon_{m} - \sum_{q}\hbar b^{\dagger}_{q}b_{q})^{-1}\nonumber\\
\Gamma(\epsilon_{k_i}) = 2\pi \sum_{m=k,l}\left|V_{am}\right|^{2}
                           \delta (\epsilon_{k_i}-\epsilon_{m} - \sum_{q}\hbar b^{\dagger}_{q}b_{q}),
\end{eqnarray}
we can rewrite the Green's operator for the resonant state as
\begin{eqnarray}
G_{aa}(\epsilon_{k_i}) =
\{[\epsilon_{k_i}-\epsilon^{0}_{a}-\sum_{q}M_{q}(b^{\dagger}_{q}+b_{-q})-\sum_{q}\hbar\omega_{q}b^{\dagger}_{q}b_{q}]
\nonumber\\
-\Delta (\epsilon_{k_i}) + \frac{i}{2}\Gamma(\epsilon_{k_i})\}^{-1}.
\end{eqnarray}
At this point, and for our present purposes, we neglect the operator nature of both $\Delta$ and $\Gamma$ as well as their
dependence on
$\epsilon$, obtaining a simplified Green's operator
\begin{equation}
G_{aa}(\epsilon_{k_i}) =
[\epsilon_{k_i}-{\bar\epsilon_{a}}-\sum_{q}M_{q}(b^{\dagger}_{q}+b_{-q})-\sum_{q}\hbar\omega_{q}b^{\dagger}_{q}b_{q}]^{-1}
\end{equation}
with a complex resonance energy ${\bar\epsilon_{a}} = \epsilon_{a}^{0}+\Delta-i\Gamma/2$, corresponding to a lorentzian
profile. 

%
%%%%%%%%%%%%%%%%%%%%%%%%%%%%%%        III. The Transition Amplitude        %%%%%%%%%%%%%%%%%%%%%%%%%%%%%%%%%%%%
%
\section{The Transition Amplitude}
Having obtained the resonance Green's operator, we now proceed to compute the transition amplitude, from the $T$ matrix
operator, $T=V+VGV$, between an initial electron-phonon state $|k_{i};n_{q}=0,\forall q\rangle$ with zero phonons and a final
state $|l_{f};n_1,n_2,\cdots,n_N\rangle$, with $n_q$ excited phonons in mode $q$, in a total of $N$ lattice vibration modes.
Since we assume no direct transitions between electron $k$ states and $l$ states, we have
\begin{eqnarray}
\langle l_{f}|T|k_{i}\rangle =& & \langle l_{f}|(\sum_{l} V_{la} c^{\dagger}_{l}c_{a})G_{aa}
(\sum_{k}V_{ak}c^{\dagger}_{a}c_{k})|k_{i}\rangle\nonumber\\
=& & V_{l_{f},a} V_{a,k_{i}} \left[\epsilon_{k_i}-{\bar\epsilon_{a}}-
\sum_{q}M_{q}(b^{\dagger}_{q}+b_{-q})-\sum_{q}\hbar\omega_{q}b^{\dagger}_{q}b_{q}\right]^{-1},
\end{eqnarray}
still a phonon operator, to be inserted between the vibrational ground state of the lattice and all possible excited states:
\begin{eqnarray}
\langle l_{f};n_{1}, n_{2}, \cdots,n_{N}|T|k_{i};0_{1},0_{2},\cdots,0_{N}\rangle
=\langle n_{1}, n_{2}, \cdots,n_{N}|\langle l_{f}|T|k_{i}\rangle |0_{1},0_{2},\cdots,0_{N}\rangle\nonumber\\
=V_{l_{f},a} V_{a,k_{i}} 
\langle n_{1}, n_{2}, \cdots,n_{N}|\left[\epsilon_{k_i}-{\bar \epsilon_{a}}- \sum_{q}M_{q}(b^{\dagger}_{q}+b_{-q})
-\sum_{q}\hbar\omega_{q}b^{\dagger}_{q}b_{q}\right]^{-1}|{\bf 0}\rangle,
\end{eqnarray}
where we wrote the lattice initial, ground state as $|{\bf 0}\rangle$.

Next, we proceed to diagonalize this last denominator by appealing to the operator $U$ (as in Refs.\ \onlinecite{D+C} and
 \onlinecite{Mahan})
\begin{equation}
U=\exp \left[-\sum_{q}\frac{M_{q}}{\hbar\omega_{q}}(b^{\dagger}_{q} - b_{-q})\right]
\end{equation}
and inserting $U^{-1} U={\bf 1}$ in the matrix element (3.2) above, after noting its effect on the phonon operators,
$b^{\dagger}_{q}$ and $b_{q}$,
\begin{eqnarray}
U b^{\dagger}_{q}U^{-1} = b^{\dagger}_{q} - \frac{M_{q}}{\hbar\omega_{q}}\nonumber\\
U b_{q}U^{-1} = b_{q} - \frac{M_{q}}{\hbar\omega_{q}},
\end{eqnarray}
results that assume symmetry of the phonon bands, specifically that $M_{q}=M_{-q}$ and $\omega_{q}=\omega_{-q}$. We have also
employed the well known operator theorem that $\exp\{A\}B\exp\{-A\}= B+[A,B]$ if $[A,B]$ is a c-number. Then, the matrix element
in (3.2) can be written
\begin{eqnarray}
\langle n_{1}, n_{2}, \cdots,n_{N}|U^{-1}U\left[\epsilon_{k_i}-{\bar \epsilon_{a}}- \sum_{q}M_{q}(b^{\dagger}_{q}+b_{-q})
-\sum_{q}\hbar\omega_{q}b^{\dagger}_{q}b_{q}\right]^{-1}U^{-1}U|{\bf 0}\rangle
\end{eqnarray}
and diagonalizes to
\begin{eqnarray}
\langle n_{1}, n_{2}, \cdots,n_{N}|U^{-1}\left[\epsilon_{k_i}-{\bar \epsilon_{a}} + \sum_{q}\frac{M^{2}_{q}}{\hbar\omega_{q}}
-\sum_{q}\hbar\omega_{q}b^{\dagger}_{q}b_{q}\right]^{-1} U|{\bf 0}\rangle.
\end{eqnarray}
We proceed to calculate this matrix element. First we note that, since $M_{q}=M_{-q}$ and $\omega_{q}=\omega_{-q}$, we can write
\begin{eqnarray}
U |{\bf 0}\rangle &=& e^{-\sum_{q}\frac{M_{q}}{\hbar\omega_{q}}(b^{\dagger}_{q}-b_{q})}|0_{1}, 0_{2},\cdots,0_{N}\rangle
\nonumber\\
&=&    e^{-\frac{M_{q_1}}{\hbar\omega_{q_1}}(b^{\dagger}_{q_1}-b_{q_1})} |0_{1}\rangle\cdot
       e^{-\frac{M_{q_2}}{\hbar\omega_{q_2}}(b^{\dagger}_{q_2}-b_{q_2})} |0_{2}\rangle\cdot\cdots\nonumber\\
& &\cdots e^{-\frac{M_{q_N}}{\hbar\omega_{q_N}}(b^{\dagger}_{q_N}-b_{q_N})} |0_{N}\rangle
\end{eqnarray}
from which, taking advantage of the operator rule $\exp\{A+B\}=\exp\{A\}\exp\{B\}\exp\{-1/2[A,B]\}$, we obtain
\begin{eqnarray}
e^{-\frac{M_{q}}{\hbar\omega_{q}}(b^{\dagger}_{q}-b_{q})}|0_{q}\rangle\nonumber\\
&=& e^{-\frac{1}{2}\left(\frac{M_{q}}{\hbar\omega_{q}}\right)^{2}}e^{-\left(\frac{M_{q}}{\hbar\omega_{q}}\right)b^{\dagger}}
e^{\left(\frac{M_{q}}{\hbar\omega_{q}}\right) b}|0_{q}\rangle \nonumber\\
&=& e^{-\frac{1}{2}\left(\frac{M_{q}}{\hbar\omega_{q}}\right)^{2}}
\sum_{m_{q}=0}^{\infty}\frac{(-M_{q}/\hbar\omega_{q})^{m_{q}}}{\sqrt{m_{q}!}}|m_{q}\rangle,
\end{eqnarray}
for each vibration mode $q$. We then obtain the product of similar terms for all $N$ phonon modes. Operating on that product
with the diagonalized denominator phonon operator, we have
\begin{eqnarray}
& &\left[\epsilon_{k_i}-{\bar \epsilon_{a}} - \sum_{q}\hbar\omega_{q}b^{\dagger}_{q}b_{q}\right]^{-1} U
|0_{1}, 0_{2},\cdots, 0_{N}\rangle\nonumber\\
&=&e^{-\frac{1}{2}\sum_{q}\left(\frac{M_{q}}{\hbar\omega_{q}}\right)^{2}}
\sum_{m_{1}=0}^{\infty}\sum_{m_{2}=0}^{\infty}\cdots\sum_{m_{N}=0}^{\infty}(-)^{\left(\sum_{q}m_{q}\right)}\nonumber\\
&\times&\frac{(M_{1}/\hbar\omega_{1})^{m_{1}}(M_{2}/\hbar\omega_{2})^{m_{2}}\cdots(M_{N}/\hbar\omega_{N})^{m_{N}}}
{\sqrt{m_{1}! m_{2}!\cdots m_{N}!}}\nonumber\\
&\times&\left[\epsilon_{k_i}-{\bar \epsilon_{a}} + \sum_{q}\frac{M_{q}^{2}}{\hbar\omega_{q}}
-\sum_{q}\hbar\omega_{q}m_{q}\right]^{-1}|m_{1},m_{2},\cdots,m_{N}\rangle.
\end{eqnarray}
It remains to calculate the action of the operator $U^{-1}$ on the intermediate vibrational state ket 
$|m_{1},m_{2},\cdots,m_{N}\rangle$:
\begin{eqnarray}
U^{-1}&&|m_{1},m_{2},\cdots,m_{N}\rangle\nonumber\\
&=&e^{\sum_{q}\frac{M_{q}}{\hbar\omega_{q}}(b^{\dagger}_{q}-b_{q})}|m_{1},m_{2},\cdots,m_{N}\rangle,
\end{eqnarray}
a product of terms of the type
\begin{eqnarray}
e^{\frac{M_{q}}{\hbar\omega_{q}}(b^{\dagger}_{q}-b_{q})}|m_{q}\rangle =&&\nonumber\\
e^{-\frac{1}{2}(\frac{M_{q}}{\hbar\omega_{q}})^{2}}\sum_{s_{q}=0}^{\infty}\sum_{r_{q}=0}^{\infty}&&(-)^{r}
\frac{(M_{q}/\hbar\omega_{q})^{r_{q}+s_{q}}}{r_{q}!s_{q}!}
\frac{\sqrt{m_{q}!(m_{q}-r_{q}+s_{q})!}}{(m_{q}-r_{q})!} |m_{q}-r_{q}+s_{q}\rangle.
\end{eqnarray}
Performing the internal product with the corresponding bra $\langle n_{q}|$, we use the orthogonality condition, 
$\langle n_{q}|m_{q}-r_{q}+s_{q}\rangle = \delta (n_{q} - m_{q} + r_{q} - s_{q})$, and get, for the $q^{th}$ mode, a
contribution (including the pre-factor in Eq.(3.8) above)
\begin{eqnarray}
&&e^{-\frac{1}{2}\left(\frac{M_{q}}{\hbar\omega_{q}}\right)^{2}}
\sum_{m_{q}=0}^{\infty}\frac{\left(-M_{q}/\hbar\omega_{q}\right)^{m_{q}}}{\sqrt{m_{q}!}}\langle n_{q}|U^{-1}|m_{q}\rangle
\nonumber\\
&&=e^{-(\frac{M_{q}}{\hbar\omega_{q}})^{2}}\left(\frac{M_{q}}{\hbar\omega_{q}}\right)^{n_{q}}(n_{q}!)^{-1/2}\nonumber\\
&&\times\sum_{m_{q}=0}^{\infty}(-)^{m_{q}}\sum_{r_{q}=0}^{m_{q}}(-)^{r_{q}}\frac{(n_{q}!)[(M_{q}/\hbar\omega_{q})^{2}]^{r_{q}}}
{r_{q}!(m_{q}-r_{q})!(n_{q}+r_{q}-m_{q})!}.
\end{eqnarray}
However, this last summation is a generalized Laguerre polynomial \cite{Lebedev,Szego,Talman}:
\begin{eqnarray}
\sum_{r_{q}=0}^{m_{q}} (-)^{r_{q}} \frac{n_{q}!}{r_{q}!(m_{q}-r_{q})!(n_{q}+r_{q}-m_{q})!}
[(M_{q}/\hbar\omega_{q})^{2}]^{r_{q}}\nonumber\\
 = L_{m_{q}}^{n_{q}-m_{q}}[(M_{q}/\hbar\omega_{q})^{2}]
\end{eqnarray}
and the whole transition amplitude matrix becomes
\begin{eqnarray}
&&\langle l_{f}; n_{1}, n_{2}, \cdots, n_{N}| T |k_{i};0_{1},0_{2},\cdots, 0_{N} \rangle\nonumber \\
&&= V_{l_{f},a} V_{a,k_{i}} e^{-\sum_{q}\left(\frac{M_{q}}{\hbar\omega_{q}}\right)^{2}} 
\left[\left(\frac{M_{1}}{\hbar\omega_{1}}\right)^{n_{1}}(n_{1}!)^{-1/2}
      \left(\frac{M_{2}}{\hbar\omega_{2}}\right)^{n_{2}}(n_{2}!)^{-1/2}\cdots
      \left(\frac{M_{N}}{\hbar\omega_{N}}\right)^{n_{N}}(n_{N}!)^{-1/2}\right]\nonumber\\
&&\times\sum_{m_{1}=0}^{\infty}\sum_{m_{2}=0}^{\infty}\cdots\sum_{m_{N}=0}^{\infty}
(-)^{m_{1}} L_{m_{1}}^{n_{1}-m_{1}}[(M_{1}/\hbar\omega_{1})^{2}]
(-)^{m_{2}} L_{m_{2}}^{n_{2}-m_{2}}[(M_{2}/\hbar\omega_{2})^{2}]
\cdots(-)^{m_{N}} L_{m_{N}}^{n_{N}-m_{N}}[(M_{N}/\hbar\omega_{N})^{2}]\nonumber\\
&&\times\left[\epsilon_{k_i}-{\bar \epsilon_{a}} + \sum_{q}\left(\frac{M_{q}^{2}}{\hbar\omega_{q}}\right) - 
(m_{1}\hbar\omega_{1} + m_{2}\hbar\omega_{2} + \cdots + m_{N}\hbar\omega_{N})\right]^{-1}
\end{eqnarray}
It is not necessary to distinguish the two cases, $n_{q}\leq m_{q}$ and $m_{q}\leq n_{q}$, if we take into consideration the
symmetry properties \cite{Szego,Talman} of the generalized Laguerre polynomials,
\begin{equation}
(-)^{\nu}\nu ! x^{-\nu}L_{\nu}^{\mu-\nu}(x) = (-)^{\mu}\mu ! x^{-\mu}L_{\mu}^{\nu-\mu}(x)
\end{equation}
which, applied to the present case, allows us to write, when $n_{q}\leq m_{q}$
\begin{eqnarray}
L_{m_{q}}^{n_{q}-m_{q}}[(M_{q}/\hbar\omega_{q})^{2}]
&=&
(-)^{n_{q}-m_{q}}\left(\frac{n_{q}!}{m_{q}!}\right)\left[\left(\frac{M_{q}}{\hbar\omega_{q}}\right)^{2}\right]^{(m_{q}-n_{q})}
\nonumber\\
& &\times L_{n_{q}}^{m_{q}-n_{q}}[(M_{q}/\hbar\omega_{q})^{2}].
\end{eqnarray}
From now on, we will always write the matrix element as displayed in Eq.(3.14) above. 
Up to this point what we have is basically the generalization, for $N$ vibration modes, of the results \cite{G+A93,D+C} for
just one vibration coordinate. But the similarities between these and the results of Wingreen et al. in Ref.\ 
\onlinecite{Wingreen} lead us to believe that the $N$-mode situation could somehow be reduced to a single vibration mode
inelastic scattering event. This is in fact possible, if we impose the restriction of an Einstein band of phonons, that is to
say $\omega_{q}=\omega_{0}$ for all modes {\bf q}. In order to perform this reduction from $N$ to just one ``mode", we begin by
considering only the last two phonon modes in Eq.(3.14) above, with the complete energy denominator:
\begin{eqnarray}
\sum_{m_{N-1}=0}^{\infty}(-)^{m_{N-1}}L_{m_{N-1}}^{n_{N-1}-m_{N-1}}(g_{N-1})
\sum_{m_{N}=0}^{\infty}(-)^{m_{N}}L_{m_{N}}^{n_{N}-m_{N}}(g_{N})\nonumber\\
\left[(z - A_{N-2}) - (m_{N-1}\hbar\omega_{N-1} + m_{N}\hbar\omega_{N})\right]^{-1}
\end{eqnarray}
where we introduced the short notations $z=\epsilon_{k_i}-{\bar \epsilon_{a}} + \sum_{q}(M_{q}^{2}/\hbar\omega_{q})$,
$A_{N-2} = m_{1}\hbar\omega_{1} + m_{2}\hbar\omega_{2} + \cdots + m_{N-2}\hbar\omega_{N-2}$ and
$g_{q}=(M_{q}/\hbar\omega_{q})^{2}$. Having isolated these factors, we keep all summation indices $m_{q}$ fixed except for
$m_{N-1}$ and $m_{N}$, impose $\omega_{N-1}=\omega_{N}=\omega_{0}$, and rewrite the double summation of these last two modes as
\begin{equation}
\sum_{m_{N-1}=0}^{\infty}\sum_{m_{N}=0}^{\infty}{\cal L}_{m_{N-1},m_{N}}
\end{equation}
which, by way of the general algebraic rule
\begin{equation}
\sum_{k=0}^{\infty}\sum_{l=0}^{\infty}{\cal L}_{k,l} = \sum_{m=0}^{\infty}\sum_{p=0}^{m}{\cal L}_{p,m-p}
\end{equation}
can now be rewritten as
\begin{equation}
\sum_{m=0}^{\infty}\frac{(-)^{m}}{(z-A_{N-2}) - m\hbar\omega_{0}}
\sum_{p=0}^{m}L_{p}^{n_{N-1}-p}(g_{N-1}) L_{m-p}^{n_{N}-m+p}(g_{N}).
\end{equation}
Appealing now to the sum rule (A7) for the generalized Laguerre polynomials, deduced in the Appendix below, the
summation over the index $p$ obeys the equality
\begin{eqnarray}
\sum_{p=0}^{m}L_{p}^{n_{N-1}-p}(g_{N-1}) L_{m-p}^{n_{N}-m+p}(g_{N})\nonumber\\
= L_{m}^{(n_{N-1} + n_{N})-m}(g_{N-1}+g_{N}),
\end{eqnarray}
and the contraction of the last two vibration modes summations originates the single summation
\begin{equation}
\sum_{m=0}^{\infty}\frac{(-)^{m}}{(z-A_{N-2}) - m\hbar\omega_{0}}L_{m}^{(n_{N-1} + n_{N})-m}(g_{N-1}+g_{N}).
\end{equation}
Keeping on contracting all Laguerre polynomials over the phonon modes, we end up with just one Laguerre polynomial and then the
transition amplitude matrix element can be written in a ``one-mode" form:
\begin{eqnarray}
\langle l_{f}; n_{1}, n_{2}, \cdots,n_{N}| T | k_{i}; 0_{1}, 0_{2},\cdots,0_{N}\rangle = \nonumber\\
V_{l_{f},a}V_{a,k_{i}} e^{-g} \left[\prod_{q}(g_{q})^{n_q/2}(n_{q}!)^{-1/2}\right]\nonumber\\
\times\sum_{m=0}^{\infty}\frac{(-)^{m}L_{m}^{n-m}(g)}{\left[\epsilon_{k_i}-{\bar \epsilon_{a}} 
+ \lambda - m\hbar\omega_{0}\right]},
\end{eqnarray}
$n=\sum_{q}n_q$ being the total number of final excited phonons into all the lattice vibration modes,
$g=\sum_{q}(M_{q}/\hbar\omega_{0})^{2}=\sum_{q}g_{q}$ and $\lambda = \sum_{q}(M_{q}^{2}/\hbar\omega_{0})$.

%
%%%%%%%%%%%%%%%%%%%%%%%%%%%%%%        IV. Summing over Final Phonon States         %%%%%%%%%%%%%%%%%%%%%%%%%%%%%%%%%%%%
%
\section{Summing over Final Phonon States}
The transition probability will be given by the modulus square of amplitude above. Summing over all possible final
phonon states $\langle n_{1}, n_{2}, \cdots,n_{N}|$ such that the total number of excited phonons is $n$, we have
\begin{eqnarray}
\left|\langle l_{f};n | T | k_{i}; {\bf 0}\rangle\right|^{2}
&=&|V_{l_{f},a}|^{2}|V_{a,k_{i}}|^{2} e^{-2g}\nonumber\\
&&\times\sum_{n_{1}} \sum_{n_{2}} \cdots \sum_{n_{N}} 
\frac{g_{1}^{n_{1}}}{n_{1}!}\frac{g_{2}^{n_{2}}}{n_{2}!}\cdots\frac{g_{N}^{n_{N}}}{n_{N}!}\nonumber\\
&&\times\left|\sum_{m=0}^{\infty}\frac{(-)^{m}L_{m}^{n-m}(g)}{\left[\epsilon_{k_i}-{\bar \epsilon_{a}} 
+ \lambda - m\hbar\omega_{0}\right]}\right|^{2}.
\end{eqnarray}
However, in general $x^{n}/n!=(-)^{n}L_{n}^{-n}(x)$ and, taking advantage once more of the algebraic properties of the
generalized Laguerre polynomials, expressed in this instance by the sum rule (A10) in the Appendix below, we obtain, keeping in
mind the restriction $n_{1}+n_{2}+\cdots+n_{N}=n$,
\begin{eqnarray}
\sum_{n_{1}} \sum_{n_{2}} \cdots&& \sum_{n_{N}}
\frac{g_{1}^{n_{1}}}{n_{1}!}\frac{g_{2}^{n_{2}}}{n_{2}!}\cdots\frac{g_{N}^{n_{N}}}{n_{N}!}\nonumber\\
=&&\sum_{n_{1}} \sum_{n_{2}} \cdots \sum_{n_{N}}
(-)^{n_{1}}L_{n_{1}}^{-n_{1}}(g_{2})(-)^{n_{2}}L_{n_{2}}^{-n_{2}}(g_{2})\nonumber\\
&&\cdots(-)^{n_{N}}L_{n_{N}}^{-n_{N}}(g_{N})\nonumber\\
=&&(-)^{n} L_{n}^{-n}(g) = \frac{g^{n}}{n!}.
\end{eqnarray}
Consequently, the transition probability from zero to all possible number $n$ of final phonons (and from electron state $k_{i}$
to electron state $l_{f}$), imposing energy conservation in the overall process, will be given by
\begin{eqnarray}
\sum_{n=0}^{\infty}\left|\langle l_{f};n | T | k_{i}; {\bf 0}\rangle\right|^{2} =&&\nonumber\\
|V_{l_{f},a}|^{2}|V_{a,k_{i}}|^{2} e^{-2g}&&\sum_{n=0}^{\infty}\frac{g^{n}}{n!}
\delta(\epsilon_{k_i}-\epsilon_{l_f}-n\hbar\omega_{0})\nonumber\\
&&\times\left|\sum_{m=0}^{\infty}\frac{(-)^{m}L_{m}^{n-m}(g)}{\left[\epsilon_{k_i}-{\bar \epsilon_{a}} 
+ \lambda - m\hbar\omega_{0}\right]}\right|^{2},
\end{eqnarray}
for all the $N$ phonon modes, but in a form that reproduces a one-mode situation, as sought. And if we now go from the domain
of wavevectors $k_{i}$ and $l_{f}$, to the domain of incoming and outgoing electron energies, $\epsilon_{i}$ and
$\epsilon_{f}$, respectively, and taking matrix elements $V_{l_{f},a}$ and $V_{a,k_{i}}$ independent of the energy in the range
of interest, we get a transmission matrix $T(\epsilon_{i},\epsilon_{f})$ given by
\begin{eqnarray}
T(\epsilon_{i},\epsilon_{f})&&\nonumber\\
&&= \Gamma_{l}\Gamma_{k}e^{-2g}\sum_{n=0}^{\infty}\frac{g^{n}}{n!}
                              \delta(\epsilon_{i}-\epsilon_{f}-n\hbar\omega_{0})\nonumber\\
&&\times\left|\sum_{m=0}^{\infty}\frac{(-)^{m}L_{m}^{n-m}(g)}{\left[\epsilon_{i}-{\bar \epsilon_{a}} 
+ \lambda - m\hbar\omega_{0}\right]}\right|^{2},
\end{eqnarray}
where $\Gamma_{k}=2\pi\sum_{k_i}|V_{a,k_{i}}|^{2}\delta(\epsilon-\epsilon_{k_i})$ and
$\Gamma_{l}=2\pi\sum_{l_f}|V_{a,l_{f}}|^{2}\delta(\epsilon-\epsilon_{l_f})$ are the partial widths of the intermediate
resonant state due to the coupling to the continuum of electron states, initial and final.

In this final result we  may represent the generalized Laguerre polynomial as
\begin{equation}
L_{m}^{n-m}(g) = \sum_{j=0}^{m} \frac{\Gamma(n+1)}{\Gamma(j+n-m+1)(m-j)!}\frac{(-g)^j}{j!}
\end{equation}
and then, using the algebraic rule in Eq.(3.19), the factor inside the modulus square of Eq.(4.3) above can be rewritten as
\begin{eqnarray}
\sum_{m=0}^{\infty}\sum_{j=0}^{m}\frac{(-)^{m}\Gamma(n+1)}{\Gamma(j-m+n+1)(m-j)!}\frac{(-g)^j}{j!}
\left[\epsilon_{i}-{\bar \epsilon_{a}} + \lambda - m\hbar\omega_{0}\right]^{-1}\nonumber\\
= \sum_{m_{1}=0}^{\infty}\sum_{m_{2}=0}^{\infty}
\frac{(-)^{(m_{1}+m_{2})}\Gamma(n+1)}{\Gamma(-m_{2}+n+1)(m_{2}!)}\frac{(-g)^{m_{1}}}{(m_{1})!}
\left[\epsilon_{i}-{\bar \epsilon_{a}} + \lambda - (m_{1}+m_{2})\hbar\omega_{0}\right]^{-1}.
\end{eqnarray}
However, if we take into consideration the presence of the poles of the gamma function $\Gamma(-m_{2}+n+1)$ at the points
$m_{2} = n+1, n+2, \cdots$, we conclude that the upper limit in the $m_{2}$ summation is, in fact, not $\infty$ but $n$.
Consequently, we can write the transmission matrix above as (changing labels, from $m_1$ to $m$, and $m_2$ to $j$)
\begin{eqnarray}
T(\epsilon_{i},\epsilon_{f}) =& &\Gamma_{l}\Gamma_{k} e^{-2g}\sum_{n=0}^{\infty}\frac{g^{n}}{n!}
                              \delta(\epsilon_{i}-\epsilon_{f}-n\hbar\omega_{0})\nonumber\\
                              & & \times\left|\sum_{m=0}^{\infty}\sum_{j=0}^{n}\frac{(-)^{j} n!}{(n-j)!j!}\frac{g^{m}}{m!}
                             \left[\epsilon_{i}-{\bar\epsilon_{a}}  + \lambda - (m+j)\hbar\omega_{0}\right]^{-1}\right|^{2},
\end{eqnarray}
which is the result expressed by Eq. (32) in Ref.\ \onlinecite{Wingreen} and by Eq. (10a) in Ref.\ \onlinecite{Gadzuk91}. We
may conclude, therefore, that the relationship between the end results of the two approaches goes beyond a numerical
indistinguishability, and that a consistent description of inelastic resonant tunneling in heterostructures can also be
achieved, without the need to use many-body Green's functions.

%%%%%%%%%%%%%%%%%%%%%%%%%%%%%%%%%%%%%%%%%        V. Conclusions           %%%%%%%%%%%%%%%%%%%%%%%%%%%%%%%%%%%%%%%%
%
\section{Conclusions}
Our aim in the present paper was to clarify the possible validity of the antecipated connection between two different
approaches to the same basic phenomenon, not only for one vibrational degree of freedom as done before, but for a whole set of
vibration modes, even if with the restriction of equal frequencies. Fundamental to the whole exposition were the
algebraic properties of the generalized Laguerre polynomials. Furthermore, we believe that it may become possible, with this
simpler method, to treat more general situations in a direct fashion, in particular the consideration of non-Lorentzian
lineshapes, of non-linear electron-phonon coupling, and of more than one intermediate resonance.

%
%%%%%%%%%%%%%%%%%%%%%%%%%%%%%%%%%%%%%%       VI.  Acknowledgments          %%%%%%%%%%%%%%%%%%%%%%%%%%%%%%%%%%%%%%%%
%
\acknowledgments

The author thanks Professor P.R. Antoniewicz for a critical reading of the manuscript and  gratefully acknowledges financial
support from F.L.A.D. (Luso-American Development Foundation) and from the University of the Azores, Portugal. 

%%%%%%%%%%%%%%%%%%%%%%%%%%%%%%%%%%%%%%%%%%            Appendix              %%%%%%%%%%%%%%%%%%%%%%%%%%%%%%%%%%%%%%%%%
%
\appendix{ }

\section*{Sum rules for the generalized Laguerre polynomials}

We give here proof of the sum rules between generalized Laguerre polynomials used above. A set of useful references 
regarding orthogonal polynomials is given below (Refs.\ \onlinecite{Lebedev,Szego,Talman,Atlas,Carlitz,Vilenkin,dutch}).
Starting from the generating function
\begin{equation}
w(x,z) = (1+z)^{\alpha} e^{- x z} = \sum_{n=0}^{\infty}c_{n}(x) z^{n} 
\end{equation}
expanded in powers of the complex variable $z$ (such that $|z|<1$), the coefficients $c_{n}(x)$ are given by
\begin{eqnarray}
c_{n}(x) =&& \frac{1}{2\pi i}\oint_{\Gamma}\frac{w(x,z)}{z^{n+1}}dz 
= \frac{1}{n!}\frac{d^{n}w(x,z)}{dz^{n}}\mid_{z=0} \nonumber\\
=&& \frac{1}{n!}\frac{d^{n}[(1+z)^{\alpha} e^{- x z}]}{dz^{n}}\mid_{z=0},
\end{eqnarray}
where the contour $\Gamma$ encloses the origin. Using Leibniz's rule for the n-th derivative of a product of two functions, we
get
\begin{eqnarray}
c_{n}(x) =& &
\frac{1}{n!}\sum_{r=0}^{n}\left(\begin{array}{c}n\\r\end{array}\right)\frac{\alpha !}{(\alpha -r)!}(-x)^{n-r}\nonumber\\
         =& &\sum_{r=0}^{n}\frac{\Gamma(\alpha+1)}{\Gamma(\alpha-n+r+1)(n-r)!}\frac{(-x)^{r}}{r!}\nonumber\\
         =& & L_{n}^{(\alpha - n)}(x).
\end{eqnarray}
Hence, we can write
\begin{equation}
w(x,z) = (1+z)^{\alpha} e^{- x z} = \sum_{n=0}^{\infty}L_{n}^{(\alpha - n)}(x) z^{n},
\end{equation}
using the definition of generalized (as opposed to ``associated") Laguerre polynomials, where the upper index can be any
complex number.
%\cite{Szego,Atlas,Carlitz}.

Some convenient representations of the generalized Laguerre polynomials, which also remain valid for $\alpha$ a negative
integer, are
%\cite{Szego,Atlas,Carlitz,dutch}
%
\begin{eqnarray}
L_{n}^{(\alpha)} =& &\sum_{r=0}^{n}\frac{\Gamma(\alpha+n+1)}{\Gamma(\alpha+r+1) (n-r)!}\frac{(-x)^{r}}{r!}\nonumber\\
=& &\frac{1}{n!} \sum_{r=0}^{n}\frac{(-n)_{r}}{r!}(\alpha+r+1)_{n-r}\/ x^{r}.
\end{eqnarray}
In this last expression we introduced the Pochhammer symbols, $(a)_{r}=\Gamma(a+r)/\Gamma(a)$.

But now, with the help of the generating function above, we can prove the sum rule required. In fact,
\begin{eqnarray}
(1+z)^{\alpha}e^{-x z}\cdot(1+z)^{\beta}e^{-y z}\nonumber\\
=& &\sum_{r=0}^{\infty}\/\sum_{s=0}^{\infty} L_{r}^{(\alpha - r)}(x) L_{s}^{(\beta - s)}(y) z^{r+s}\nonumber\\
=& &(1+z)^{\alpha + \beta} e^{-(x+y) z}\nonumber\\
=& & \sum_{n=0}^{\infty} L_{n}^{(\alpha+\beta-n)}(x+y) z^{n}
\end{eqnarray}
Equating equal powers of the variable $z$, we conclude that $r+s=n$ and consequently
\begin{equation}
\sum_{r=0}^{n} L_{r}^{(\alpha - r)}(x) L_{n-r}^{(\beta + r - n)}(y) = L_{n}^{(\alpha+\beta-n)}(x+y),
\end{equation}
our first desired result. The summation index $r$ can not attain values larger than $n$, since the lower index of a generalized
Laguerre polynomial is always nonnegative.

Next we prove the second sum rule used above. It is, in fact, a consequence of the first sum rule, for which the
particular case  $\alpha=\beta=0$ gives
\begin{equation}
\sum_{r=0}^{n} L_{r}^{ (- r)}(x_{1}) L_{n-r}^{(r - n)}(x_{2}) = L_{n}^{(-n)}(x_{1}+x_{2}).
\end{equation}
If we now consider only the first two factors in the product (4.2) above, we may write, for a fixed value of $n$ (and of 
$n_3,\cdots,n_N$),
\begin{eqnarray}
\sum_{n_{1}+n_{2}=n-(n_{3}+\cdots+n_{N})} && L_{n_{1}}^{-n_{1}}(g_{1}) L_{n_{2}}^{-n_{2}}(g_{2})\nonumber\\
= && \sum_{n_{1}=0}^{n-(n_{3}+\cdots+n_{N})} L_{n_{1}}^{-n_{1}}(g_{1}) 
L_{[n-(n_{3}+\cdots+n_{N})-n_{1}]}^{-[n-(n_{3}+\cdots+n_{N})-n_{1}]}(g_{2})\nonumber\\
= && L_{[n-(n_{3}+\cdots+n_{N})]}^{-[n-(n_{3}+\cdots+n_{N})]}(g_{1}+g_{2}).
\end{eqnarray}
We repeat this contracting process over all $N$ factors, ending up with just one final Laguerre polynomial, $ L_{n}^{-n}(g)$. In
general, then,
\begin{eqnarray}
\sum_{n_{1}+n_{2}+\cdots+n_{N} = n}&& L_{n_{1}}^{-n_{1}}(x_{1}) L_{n_{2}}^{-n_{2}}(x_{2})\cdots
L_{n_{N}}^{-n_{N}}(x_{N})\nonumber\\
&& = L_{n}^{-n}(x),
\end{eqnarray}
with $x = x_{1}+x_{2}+\cdots+x_{N}$ and $n = n_{1}+n_{2}+\cdots+n_{N}$.

%%%%%%%%%%%%%%%%%%%%%%%%%%%%%%%%%%%%%%%%%        REFERENCES      %%%%%%%%%%%%%%%%%%%%%%%%%%%%%%%%%%%%%%%%%%%%%%%%%%

%
%
\end{document}